\documentstyle[fleqn]{tp}

\input psfig

\def\Npix{{N_{\rm pix}}}

\begin{document}

\twocolumn[
\title{Constraints on Cosmological Parameters from Current CMB Data}
\author{James G. Bartlett, Alain Blanchard, Marian Douspis, Morvan Le Dour\\
{\it Observatoire de Strasbourg, 67000 Strasbourg, FRANCE}\\}
\vspace*{16pt}   

ABSTRACT.\
We discuss the constraints one can place on cosmological parameters
using current cosmic microwave background data.  A standard
$\chi^2$--minimization over band--power estimates is first
presented, followed by a discussion of the more correct 
likelihood approach.  We propose an approximation to the 
complete likelihood function of an arbitrary experiment
requiring only limited and easily found information about
the observations.  Examination of both open models --
$(\Omega,h,Q,n)$ -- and flat models ($\Omega+\Omega_\Lambda=1$) 
-- $(\Omega,\Omega_b,h,Q,n)$ -- leaves one rather robust result:
models with small curvature are favored.
\endabstract]

\markboth{Bartlett et al.}{Constraints on Cosmological Parameters...}

\small

\begin{figure*}
\centering\mbox{\psfig{figure=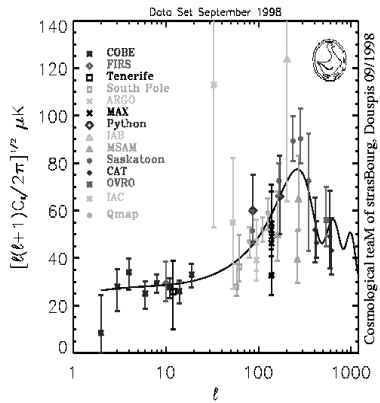,height=6cm}}
\caption[]{Current CMB power spectrum estimates.  The curve
shows the predictions of a model with $\Omega=0.9$,
$h=0.35$, $\Omega_bh^2=0.015$, $Q=17\mu$K and $n=0.94$.}
\label{fig:1*}
\end{figure*}

\section{Introduction}
Current cosmic microwave background (CMB) results
already permit some constraints to be placed on certain 
cosmological parameters, such as the density parameter, 
$\Omega$, the Hubble constant, $H_o \equiv h (100$ km/s/Mpc), etc... 
(Bond \& Jaffe 1996; Lineweaver et al. 1997;  
Hancock et al. 1998; Bartlett et al. 1998a).  
The present data set is most 
clearly summarized in the power spectrum plane as a 
set of (flat) band--power estimates, as shown in
Figure 1.  Presence of the ``Doppler peak'' is 
indicated by several observations on its upward
slope, including the new QMAP results 
(Oliveira--Costa et al. 1998), and most importantly
the Saskatoon points (Netterfield et al. 1997).  
It is worth emphasizing, all 
the same, that MSAM (Cheng et al. 1997) is not as supportive of the peak 
as Saskatoon, and these are the only two experiments
covering the peak at the present time.  Analysis of 
newly obtained data, e.g., from BOOMERANG'S North American 
flight, MAXIMA or MAT, all capable of measuring the
peak with good precision, will help to resolve the issue.

	We have used these power
spectrum estimates to constrain cosmological
parameters in both open ($\Omega_\Lambda=0$) and flat 
($\Omega+\Omega_\Lambda=1$) scenarios 
within the context of inflationary models
with cold dark matter.  After presenting some of our results
based on a $\chi^2$--minimization, we discuss the
shortcomings of this approach.  An easy--to--use
approximation to the full likelihood function of 
an experiment is then proposed and appears to lead
to less restrictive constraints than indicated by 
the $\chi^2$--minimization.  Nevertheless,
one result remains the same, despite the general
differences between the two methods: purely open 
models with $\Omega<0.4$ are strongly disfavored,
arguing against a Universe with large curvature. 

\section{$\chi^2$--approach}

\begin{figure*}
\centering\mbox{\psfig{figure=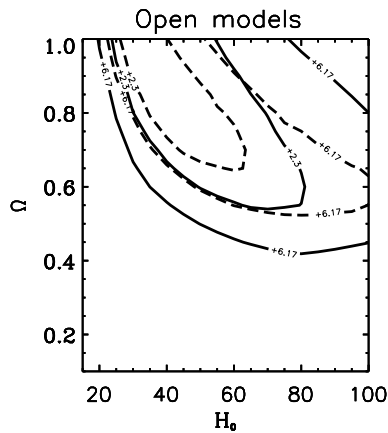,height=6cm}}
\caption[]{Open model constraints.  The dashed lines
show the $\chi^2$--minimization constraints (68\% and
95\% confidence boundarys), while
the solid lines correspond to the approximate likelihood
function.}
\label{fig:2*}
\end{figure*}

\begin{figure*}
\centering\mbox{\psfig{figure=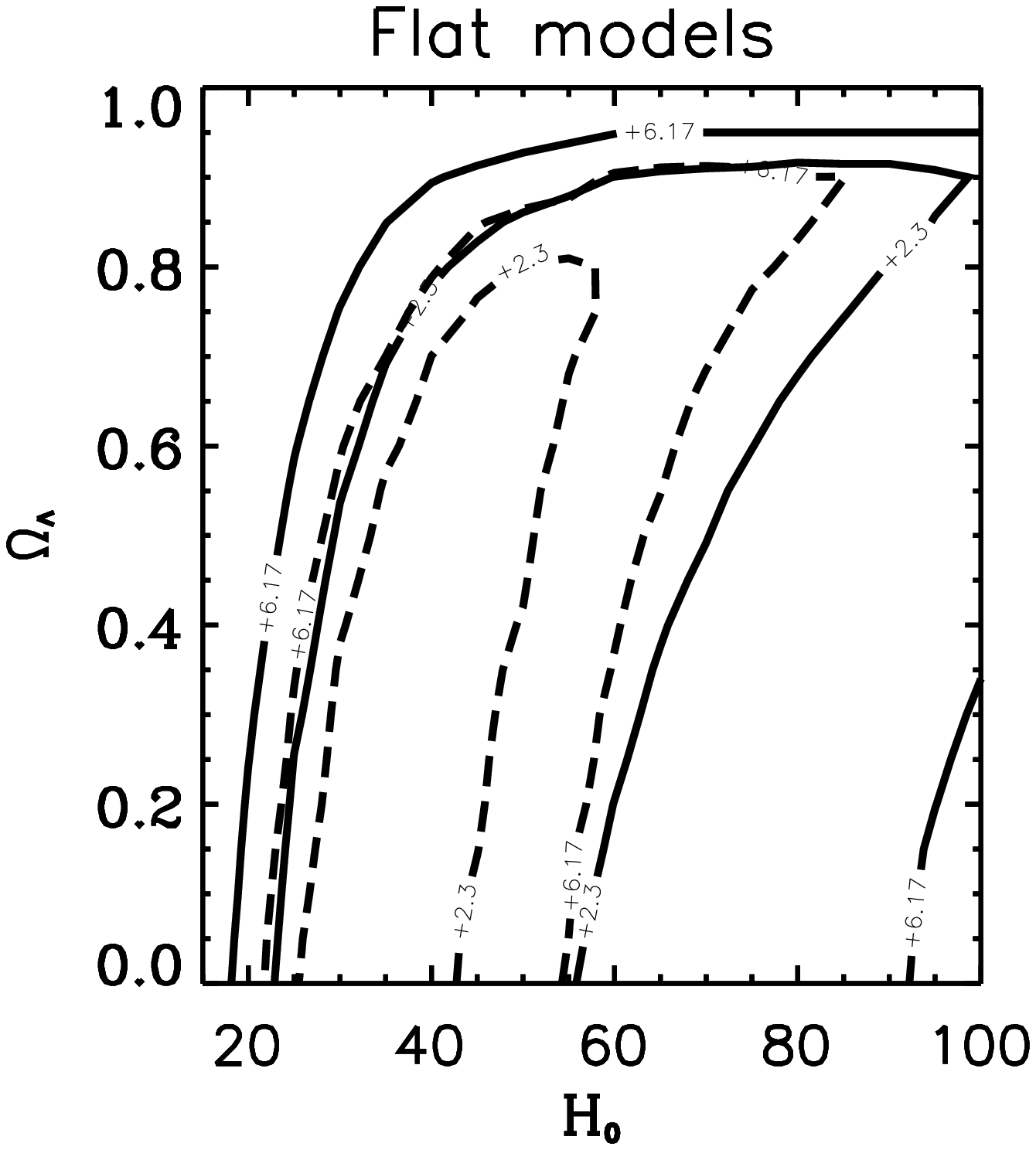,height=6cm}}
\caption[]{Flat model constraints.  The lines have the
same meaning as in Figure 2.}
\label{fig:3*}
\end{figure*}

	The simplest approach is a straight--forward 
application of the $\chi^2$--statistic to the data shown
in Figure 1.  We used CMBFAST (Zaldarriaga et al. 1998;
Seljak \& Zaldarriaga 1996) to construct a grid of 
power spectra spanned by $(\Omega, h, Q, n)$, for purely open 
models, and by $(\Omega, \Omega_b, h, Q, n)$ for  
flat models with $\Omega+\Omega_\Lambda=1$.  
In Figures 2 and 3 we show the resulting 
constraints on $\Omega$ and $h$ as dashed lines 
for both types of models.
The confidence regions are defined relative to 
the minimum value of $\chi^2$; for example, 
over two parameters the 68\% and 95\% contours
are defined by the projection onto the plane
of the multidimensional ellipse inscribed by 
$\chi^2_{min} + 2.3$ and $\chi^2_{min} + 6.17$,
respectively.
at
	The most striking result is the lower limit on 
the density parameter in open models -- $\Omega>0.5$ at
95\% confidence.  This is quite clearly due to 
the apparent position of the Doppler peak around $l=300$ 
in the data.  In this sense, the current data do not
favor large spatial curvature, which would move the peak
too far to the right in the figure.  An illustrative 
example of this is given in Figure 4, where we see 
how miserably a somewhat ``conservative'' model
with $\Omega=0.2$ and $h=0.6$ performs.  
 
\section{Of Likelihood Functions}
	
	The common method just employed does not strictly apply
to the CMB data in Figure 1, for it supposes that the
data points are gaussian distributed, and this is 
certainly {\bf not} the case.  Even if the temperature
of individual pixels on the sky can be considered 
a gaussian random variable,
as predicted by inflationary scenarios, if the
experimental noise is also gaussian, a band--power estimate
will not be gaussian, because it represents the {\em variance}
of the pixel fluctuations.  The variance of a gaussian random
variable is not itself gaussian distributed.  Thus, at a 
fundamental level, the $\chi^2$--approach is not suited to
our problem.  

	We must employ a more general likelihood 
function to properly constrain our parameters.  
This may be done for a given experiment if one
assumes that the temperature fluctuations on the sky, the
pixel values, are indeed gaussian random variables.  
In this case, with a set of $\Npix$
pixel values arranged in a data vector $\vec{d}$, we
may write the likelihood function for the parameters,
represented by the vector $\vec{\Theta} = \{\Omega, h,...\}$, as
\begin{eqnarray}
{\cal L}(\vec{\Theta}) & \equiv & Prob(\vec{d}|\vec{\Theta}) \\
\nonumber
& = & \frac{1}{(2\pi)^{\Npix/2} |{\bf C}|^{1/2}} 
	e^{-\frac{1}{2}\vec{d}^t \cdot {\bf C}^{-1} \cdot \vec{d}} 
\end{eqnarray}
The key object is the covariance matrix of the pixels, ${\bf C}$:
\begin{eqnarray}
C_{ij} & \equiv & <d_id_j> = C(\theta_{ij}) + N_{ij} \\
\nonumber
& & = \frac{1}{4\pi}
	\sum_l (2l+1) C_l [\vec{\Theta}] |W_l|^2 \\
\nonumber
& & \times P_l(\cos\theta_{ij}) + N_{ij}
\end{eqnarray}
where $W_l$ is the window function defined by the experimental
beam and $N_{ij}$ is the experimental noise covariance matrix.
We have imagined a situation were the observations
are represented by a set of simple pixel values, e.g., a map, 
but the argument is the same for temperature differences,
or any linear combination of pixel values.  

	The correct manner to proceed would be to
obtain ${\cal L}$ for each experiment (including all 
correlations between experiments covering the same sky
zone) and form the complete likelihood function for the
entire data set.  This is rather demanding and requires 
sometimes hard to find experimental information, such
as a full noise covariance matrix.  We have
developed a general purpose approximation to
the likelihood function which, by comparison
to the full likelihood functions from COBE, Saskatoon 
and MAX, seems to work well (work in progress, and reported in 
Bartlett et al. 1998b; for other work, see Bond et al. 1998
and Wandelt et al. 1998).  The approximation requires
only limited information concerning a given 
experiment, such as the number of pixels, the
band--power estimate and error bars, and an idea
of the noise level -- all readily found in the literature.

\begin{figure*}
\centering\mbox{\psfig{figure=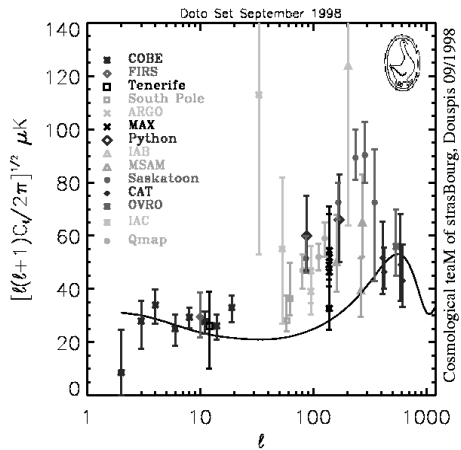,height=6cm}}
\caption[]{A model with $\Omega=0.2$, $H_o=60$ km/s/Mpc, $n=1$,
$Q=20\; \mu$K, $\Omega_bh^2 = 0.015$}
\label{fig:4}
\end{figure*}

\section{Results from the Approximate Likelihood}

	Applying our approximate likelihood ansatz to each
data point in Figure 1, we have revisited the parameter
constraints imposed by the data set.  The results in
the $(\Omega,h)$--plane are shown in Figures 2 and 3
as the solid lines; once again, these are projections
of ellipses defined relative to the maximum value 
of the likelihood. We see that
in general the confidence regions are larger 
than those deduced from the $\chi^2$--approach.  
This is most spectacular for the flat models, where
virtually all of the parameter space in this plane
is now allowed at ``95\%'' confidence. 
One should be cautious interpreting these
results, however, for a more correct definition of 
confidence regions would integrate the 
likelihood function over the parameters not
shown.  In any case, the difference
in the contours, defined in the same way,
clearly indicates that the approximate likelihood 
function has a different distribution over the
parameter space than that suggested by the
$\chi^2$--approach.    

	Thus, we see that care must be taken in drawing 
conclusions from a $\chi^2$--minimization
to band--power estimates.  Nevertheless, the best fit models
do not change drastically.  We also find about the same
lower limit as before to $\Omega$ in open models --
$\Omega>0.4$.

\section{Conclusion}

	It is important to realize that the CMB 
is already beginning to divulge its treasures
of information.  The method used to extract
quantitative constraints on cosmological 
parameters requires some care, and a
simple $\chi^2$--analysis using band--power 
estimates is not the best approach, leading to constraints
which appear tighter than the data warrant.  One
important result already indicated by the present data
is that the spatial curvature of the Universe cannot
be too large; this is quantified by the lower limit
of $\Omega>0.4$ for the open models examined here.



\end{document}